# Anomalous Threshold Reduction from <100> Uniaxial Strain for a Low-Threshold Ge Laser

David S. Sukhdeo[1], Shashank Gupta[1], Krishna C. Saraswat[1], Birendra (Raj) Dutt[2,3], and Donguk Nam[*,4]

[1]Department of Electrical Engineering, Stanford University, Stanford, CA 94305, USA
[2]APIC Corporation, Culver City, CA 90230, USA
[3]PhotonIC Corporation, Culver City, CA 90230, USA
[4]Department of Electronics Engineering, Inha University, Incheon 402-751, South Korea
[*]Email: dwnam@inha.ac.kr

**Abstract:** We theoretically investigate the effect of <100> uniaxial strain on a Ge-on-Si laser using deformation potentials. We predict a sudden and dramatic ~200x threshold reduction upon applying sufficient uniaxial tensile strain to the Ge gain medium. This anomalous reduction is accompanied by an abrupt jump in the emission wavelength and is explained by how the light-hole band raises relative to the heavy-hole band due to uniaxial strain. Approximately 3.2% uniaxial strain is required to achieve this anomalous threshold reduction for $1 \times 10^{19}$ cm$^{-3}$ n-type doping, and a complex interaction between uniaxial strain and n-type doping is observed. This anomalous threshold reduction represents a substantial performance advantage for uniaxially strained Ge lasers relative to other forms of Ge band engineering such as biaxial strain or tin alloying. Achieving this critical combination of uniaxial strain and doping for the anomalous threshold reduction is dramatically more relevant to practical devices than realizing a direct band gap.

## *Introduction*

Strain engineering has been traditionally important in semiconductor device technology because of its ability to substantially modify a material's bandstructure and thereby change its electrical and optical properties in useful ways [1]–[5]. Tensile strain engineering in germanium (Ge) has been an especially major research topic within the silicon- (Si-) compatible photonics community [4], [6]–[8]. Ge can be readily integrated on Si [9] and, upon application of sufficient tensile strain, it has the potential to become an efficient gain medium for a CMOS-compatible light source [8], [10]–[12], with possible applications in both on-chip optical interconnects [4], [6], [7], [13], [14] and infrared sensing [15]–[18]. Many theoretical analyses of biaxially strained Ge lasers exist in literature [8], [10]–[12], but modeling of uniaxially stained Ge lasers remains much more limited. This is a serious deficiency in light of recent experimental advances in achieving large, spatially homogeneous CMOS-compatible <100> uniaxial strains in Ge [14], [19], [20], including up to 5.7% uniaxial strain which is sufficient to turn Ge into a direct bandgap semiconductor [21].

Here we present extensive theoretical modeling of the threshold current density in <100> uniaxially strained Ge lasers. One of the most important and surprising features we predict is the sudden reduction of threshold current by more than two orders of magnitude on application of uniaxial tensile strain in the range of 2-4%. The exact amount of critical strain needed for this dramatic reduction depends on the amount of n-type doping in the Ge, and the anomalous threshold reduction itself is due to valence band splitting. Under biaxial tensile strain, the light-hole and heavy-hole bands split in such a way that the top valence band becomes heavy-hole like [3], [22], [23] and, if this splitting is large enough, there can be a sudden change in lasing behavior [8]. This is because both light-hole and heavy-hole bands contribute to the optical gain when the valence band splitting is small, but when the splitting is large only the top valence band will contribute to the gain[8]. For uniaxial strain, however, this valence band splitting is even more beneficial because it causes the top valence band to be the light-hole band, thus dramatically reducing the density of states at the top valence band [3], [22], [23]. This particularity of uniaxial strain has been somewhat overlooked by researchers despite its immense importance. Here we discuss this valence band splitting effect of uniaxial strain for reducing Ge laser threshold and show the importance of properly choosing the design parameters of n-type doping and uniaxial strain in order to realize a truly practical Ge laser with a low threshold.

Uniaxial strain changes the band structure in qualitatively the same way as biaxial strain, but with the valence band splitting inverted. We can directly model these changes to Ge's bandstructure under <100> uniaxial strain using deformation potential theory [3] as shown in Fig. 1. Whereas it takes 1.7% biaxial strain to achieve a direct bandgap, we find that 4.6% uniaxial strain is required to accomplish the same feat [24]. The major difference between uniaxial and biaxial strain concerns the valence band splitting that occurs under strain. Whereas for <100> biaxial tensile strain the top valence band retains a mostly heavy-hole (HH) character, under <100> uniaxial strain the LH/HH splitting is such that the top valence band retains a mostly light-hole (LH) character [22], [23]. We will see through laser modeling how this distinction concerning valence band splitting will play a very major role in reducing the threshold with uniaxial strain.

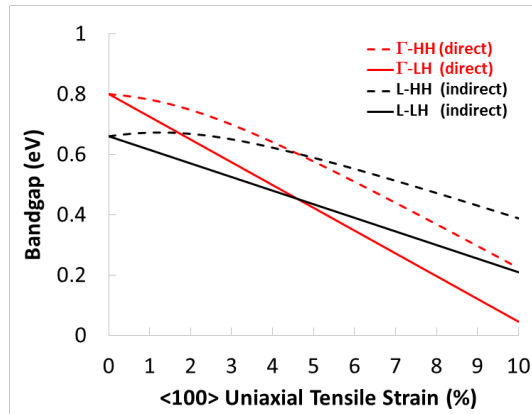

***Fig. 1.*** *Ge's direct (Γ) and indirect (L) bandgap energies vs. <100> uniaxial tensile strain.*

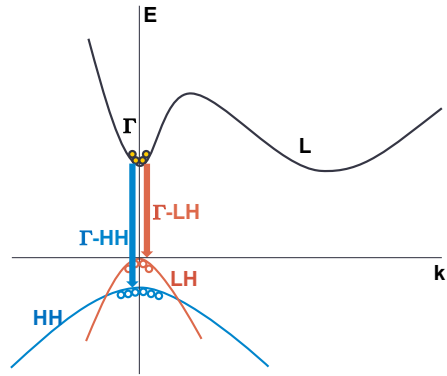

***Fig. 2.*** *Illustration of the distinct Γ-LH and Γ-HH transitions, respectively associated with the light-hole (LH) and heavy-hole (HH) bands, which can both contribute to optical gain.*

Having determined precisely how the band edges change with uniaxial strain, we can now compute the threshold current density as a function of uniaxial strain and n-type doping following the method we employed previously in Ref. [8] . For every strain value, we calculate the quasi-Fermi levels as a function of carrier densities assuming Fermi-Dirac carrier statistics and the band edges of Fig. 1. Then, we compute the net optical gain using the strain-dependent absorption coefficient of Ref. [25] and the empirical free carrier absorption relation of Ref. [1], and determine the steady-state injected carrier density at the threshold condition (i.e. when the net optical gain equals to the assumed resonator loss which is assumed to be zero in this work). Finally we obtain the threshold current by using Ge's recombination coefficients [1]. The use of the strain-dependent absorption coefficient of Ref. [25] allows us to properly treat the separate contributions of the two allowed interband transitions, Γ-LH and Γ-HH, which are indicated in Fig. 2. In this modeling we presume a double heterostructure design with perfect carrier confinement and a 300nm-thick Ge active region that always lases at the wavelength of maximum net gain [8], [26] .

Valence band splitting, illustrated in Fig. 2, occurs under any non-hydrostatic strain and will prove particularly important for our analysis. For <100> biaxial tensile strain, the valence band splitting raises the heavy-hole band relative to the light-hole band [3], [8], [22]. For the biaxial case this valence band splitting was shown to be helpful [27], but still the density of states at the top of the valence band remained quite high as the heavy-hole band was on top [8], [22]. In contrast, under <100> uniaxial tensile strain the top valence band is the light-hole band, which results in a dramatically reduced density of states at the top of the valence band [3], [22]. This plays a major role in determining the threshold current density and preferred emission wavelength as a function of uniaxial strain and n-type doping, computed using the previously described method and shown in Fig. 3.

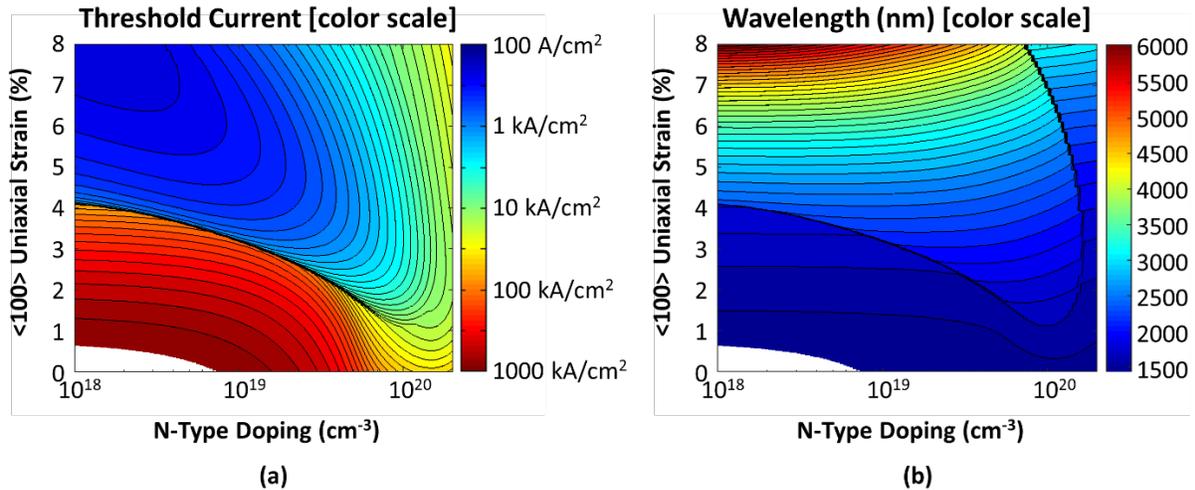

*Fig. 3.* *(a) Threshold current density (color scale) of a double heterostructure Ge laser vs. uniaxial tensile strain and n-type doping. (b) Emission wavelength (color scale) vs. uniaxial tensile strain and n-type doping. Ge thickness is assumed to be 300nm with an optical cavity loss of zero and a defect-limited minority carrier lifetime of 100ns. The blank region in the bottom left corner is due to the cutoff of the simulation bounds, i.e. thresholds greater than 1000 kA/cm$^3$.*

From Fig. 3, we observe many similarities when compared to the <100> biaxial strain scenario which was modeled in Ref. [8]. Both uniaxial and biaxial strain, for instance, exponentially reduce the threshold current of a Ge laser and dramatically redshift the preferred emission wavelength. These enhancements with strain continue even after a direct bandgap has been achieved at 4.6% uniaxial strain as we show here in Fig. 3(a) and indicates that there is no special benefit to achieving a direct bandgap with respect to laser performance, something we have previously shown for biaxial strain too [8]. We can also observe two distinct regimes for both uniaxial and biaxial strain due to LH/HH splitting in the valence bands: after about 2-4% uniaxial strain it becomes possible to achieve particularly dramatic reductions in threshold coupled with an abrupt jump in wavelength. This two-regime phenomenon is where uniaxial strain becomes substantially more promising than biaxial strain. For uniaxial strain we find that exceeding a certain critical value causes an abrupt and immediate drop in the threshold itself while for biaxial strain we did not find such an abrupt drop in the threshold but instead simply a kink in the threshold response [8]. This phenomenon with uniaxial strain can be seen more clearly in Fig. 4, which is a 3D surface plot of Fig. 3(a). This is a very important distinction between uniaxial and biaxial strain, since this sudden drop in threshold under uniaxial strain can exceed two orders of magnitude depending on the doping level.

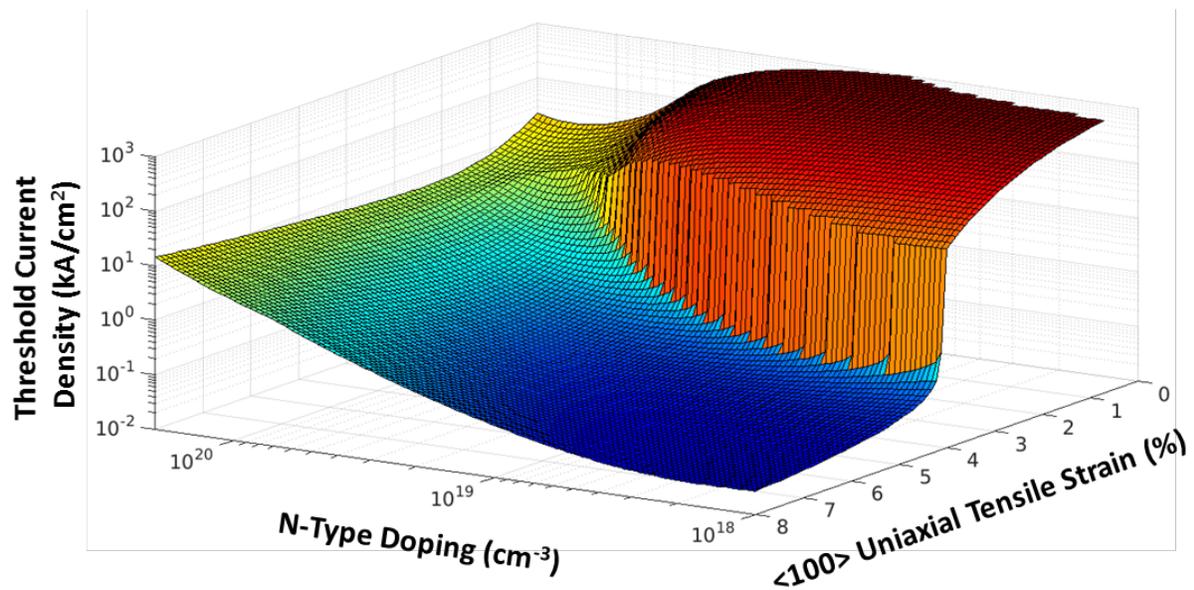

**Fig. 4.** *Threshold current density vs. uniaxial tensile strain (<100> orientation) and n-type doping. A double heterostructure design with a 300nm-thick Ge active region and zero optical cavity loss is assumed.*

The explanation for why only uniaxial strain allows for this sudden and abrupt drop in threshold at a critical value has to do with the details of the LH/HH splitting within the valence bands. For biaxial strain the top valence band retains a mostly heavy-hole (HH) character which, due to its larger density of states, means that there is less immediate benefit when the splitting is sufficiently large to allow lasing from just top valence band transition. For uniaxial strain, on the other hand, the top valence band retains a mostly light-hole (LH) character. This means that when the valence band splitting does become sufficiently large to allow lasing using only the Γ-LH transition, the threshold will be dramatically smaller due to the much smaller density of states in the LH band. In addition, due to the smaller joint density of states for the Γ-LH transition, a large amount of valence band splitting (and thus a large amount of uniaxial strain) is required before the gain from this Γ-LH transition is alone sufficient to overcome free carrier absorption. This is why a comparatively large 2-4% uniaxial strain is required to achieve disproportionate threshold reductions.

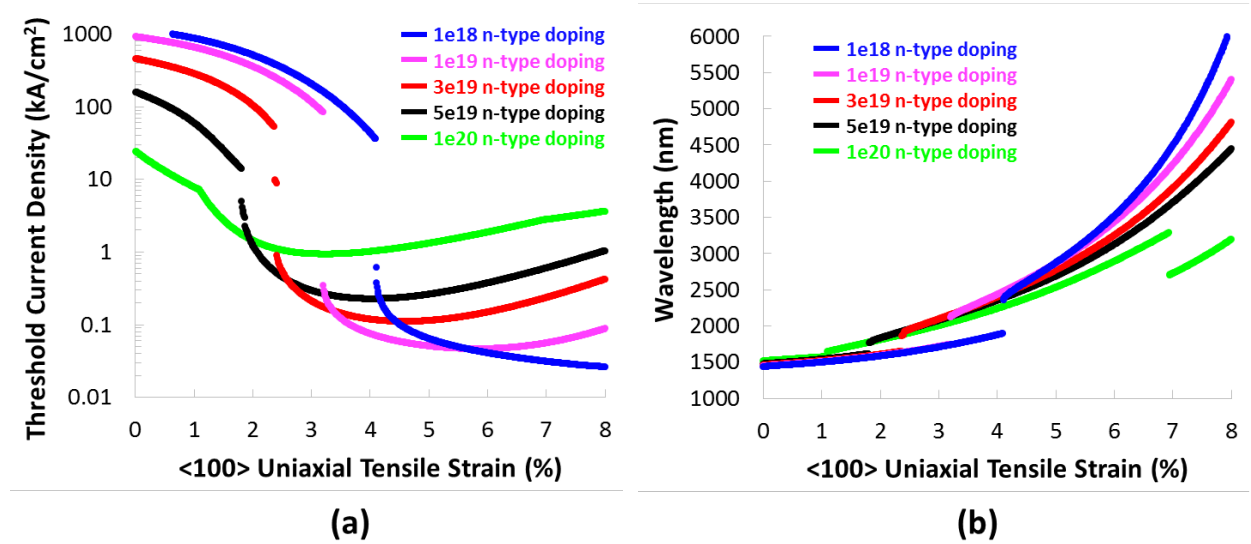

*Fig. 5. (a) Threshold current density of a Ge laser vs. uniaxial tensile strain (<100> orientation) for different doping conditions. (b) Emission wavelength of a Ge laser vs. uniaxial tensile strain (<100> orientation) for different doping conditions. In all cases a double heterostructure design with a 300nm thick Ge active region, zero optical cavity loss, and a defect-limited minority carrier lifetime of $\tau_{SRH}$=100ns is assumed. The outlier data points in the discontinuity region of figure (a) are artifacts from linear interpolations in the vicinity of the discontinuity.*

As shown in Fig. 5, employing n-type doping (or increasing the n-type doping level) allows this regime change to happen at smaller strain values whereupon the discontinuity is less pronounced. At $1 \times 10^{20}$ cm$^{-3}$ n-type doping, the regime change occurs at ~1.1% strain and there is no discontinuity, just a kink. At $5 \times 10^{19}$ cm$^{-3}$ doping, the regime change occurs at ~1.8% strain and there is a discontinuity whereupon the threshold abruptly drops by a factor of ~5x. At $1 \times 10^{19}$ cm$^{-3}$ doping, the regime change occurs at ~3.2% strain and there is a discontinuity whereupon the threshold abruptly drops by a factor of ~200x. This makes n-type doping a double-edged sword with respect to improving uniaxially strained Ge lasers. On one hand, applying n-type doping makes the critical strain smaller and thus easier to reach, but on the other hand n-type doping reduces the benefits that can be realized by exceeding this critical strain. This fits with the general theme of moderate n-type doping being useful to strained Ge laser performance, but too much n-type doping proving to be quite harmful [8].

This means that it is truly imperative to reach this critical strain level for uniaxial strain if a low-threshold Ge laser is to be realized. Whereas for the biaxial case this critical strain represents a soft boundary where the threshold shows faster improvements beyond a certain strain [8], for the uniaxial case this critical strain represents a hard boundary between a very high threshold regime and a very low threshold regime. For the case of $1 \times 10^{19}$ cm$^{-3}$ n-type doping, this critical strain is about 3.2% along the <100> direction which is considerably smaller than the 4.6% <100> uniaxial strain needed to achieve a direct bandgap. What this means is that, in order to dramatically reduce the lasing threshold, it is much more important to properly choose design parameters such as n-type doping and uniaxial strain level than to simply pursue the 4.6% uniaxial strain for achieving the direct bandgap Ge which has been the central objective of many researchers.

## *Summary*

In summary, we have performed in-depth modeling of <100> uniaxial strain, with a particular focus on the effect of valence band splitting. This modeling suggests that uniaxial strain can be much more effective than biaxial strain with respect to improving the performance of Ge lasers. This special benefit from uniaxial strain comes from the fact that under uniaxial tensile strain the top valence band retains a mostly light-hole character whereas under biaxial tensile strain the top valence band retains a mostly heavy-hole character. This means that uniaxial strain is dramatically more effective than biaxial strain at reducing the density of states at the top of the valence band. As a result, we find that the threshold drops even more dramatically with uniaxial strain than with biaxial strain. With uniaxial strain there exists a discontinuity in the threshold versus strain curve such that the threshold abruptly drops when the strain exceeds a critical value (2-4% depending on the n-type doping level). Thus, while our results indicate that there is no practical utility in striving for a direct bandgap at 4.6% uniaxial strain, there is immense utility in exceeding the easier goal of ~4% uniaxial strain which unlocks the anomalous threshold reduction in undoped Ge. With $5 \times 10^{19}$ cm$^{-3}$ n-type doping, a doping level which is practical [28], the critical uniaxial strain for which researchers must strive becomes as low as 1.8%. This is considerably lower than what has been reported in stressed micro-bridge structures [19]–[21], and even techniques incorporating an optical cavity now report strains very close to this value [29]. This indicates that practical realization of a low threshold uniaxially strained Ge laser may be well on the way to fruition if recent advances in strain engineering [19]–[21] [29] can be combined with effective device integration.

## *Acknowledgements*

This work was supported by the Office of Naval Research (grant N00421-03-9-0002) through APIC Corporation (Dr. Raj Dutt) and by a Stanford Graduate Fellowship. This work was also supported by an INHA UNIVERSITY Research Grant and by the Pioneer Research Center Program through the National Research Foundation of Korea funded by the Ministry of Science, ICT & Future Planning (2014M3C1A3052580). The authors thank Boris M. Vulovic of APIC Corporation for helpful discussions.